\documentclass[aps, prb, twocolumn, reprint, superscriptaddress, floatfix, citeautoscript, longbibliography]{revtex4-1}
\usepackage[T1]{fontenc} 
\usepackage[utf8]{inputenc}
\usepackage{newtxtext}
\usepackage{newtxmath}
\usepackage{graphicx}
\usepackage{siunitx}
\usepackage{booktabs}
\usepackage[english]{babel}
\usepackage{csquotes}
\usepackage{microtype}
\usepackage[unicode,colorlinks=true,allcolors=blue]{hyperref}
\usepackage{mathrsfs}

\usepackage{mathtools, nccmath}

\usepackage{cleveref}

\newcommand{\bnmr}{$\beta$-NMR}

\newcommand{\msr}{$\mu$SR}

\newcommand{\eli}{\textsuperscript{8}{Li}}
\newcommand{\elip}{\textsuperscript{8}{Li}\textsuperscript{+}}

\newcommand{\lip}{{Li}\textsuperscript{+}}

\newcommand{\rno}{$R$NiO$_{3}$}
\newcommand{\lno}{LaNiO$_{3}$}

\begin{document} 

\title{Local Metallic and Structural Properties of the Strongly Correlated Metal LaNiO$_3$ using $^8$Li $\beta$-NMR}

\author{Victoria L. Karner}
\email[Email: ]{vkarner@chem.ubc.ca}
\affiliation{Department of Chemistry, University of British Columbia, Vancouver, BC V6T~1Z1, Canada}
\affiliation{Stewart Blusson Quantum Matter Institute, University of British Columbia, Vancouver, BC V6T~1Z4, Canada}

\author{Aris Chatzichristos}
\affiliation{Stewart Blusson Quantum Matter Institute, University of British Columbia, Vancouver, BC V6T~1Z4, Canada}
\affiliation{Department of Physics and Astronomy, University of British Columbia, Vancouver, BC V6T~1Z1, Canada}

\author{David L. Cortie}
\altaffiliation{Current address: Institute for Superconducting and Electronic Materials, Australian Institute for Innovative Materials, University of Wollongong, North Wollongong, NSW 2500, Australia}
\affiliation{Department of Chemistry, University of British Columbia, Vancouver, BC V6T~1Z1, Canada}
\affiliation{Stewart Blusson Quantum Matter Institute, University of British Columbia, Vancouver, BC V6T~1Z4, Canada}
\affiliation{Department of Physics and Astronomy, University of British Columbia, Vancouver, BC V6T~1Z1, Canada}

\author{Martin H. Dehn}
\affiliation{Stewart Blusson Quantum Matter Institute, University of British Columbia, Vancouver, BC V6T~1Z4, Canada}
\affiliation{Department of Physics and Astronomy, University of British Columbia, Vancouver, BC V6T~1Z1, Canada}

\author{Oleksandr Foyevtsov}
\affiliation{Stewart Blusson Quantum Matter Institute, University of British Columbia, Vancouver, BC V6T~1Z4, Canada}
\affiliation{Department of Physics and Astronomy, University of British Columbia, Vancouver, BC V6T~1Z1, Canada}

\author{Kateryna Foyevtsova}
\affiliation{Stewart Blusson Quantum Matter Institute, University of British Columbia, Vancouver, BC V6T~1Z4, Canada}
\affiliation{Department of Physics and Astronomy, University of British Columbia, Vancouver, BC V6T~1Z1, Canada}

\author{Derek Fujimoto}
\affiliation{Stewart Blusson Quantum Matter Institute, University of British Columbia, Vancouver, BC V6T~1Z4, Canada}
\affiliation{Department of Physics and Astronomy, University of British Columbia, Vancouver, BC V6T~1Z1, Canada}

\author{Robert F. Kiefl}
\affiliation{Stewart Blusson Quantum Matter Institute, University of British Columbia, Vancouver, BC V6T~1Z4, Canada}
\affiliation{Department of Physics and Astronomy, University of British Columbia, Vancouver, BC V6T~1Z1, Canada}
\affiliation{TRIUMF, 4004 Wesbrook Mall, Vancouver, BC V6T~2A3, Canada}

\author{C. D. Philip Levy}
\affiliation{TRIUMF, 4004 Wesbrook Mall, Vancouver, BC V6T~2A3, Canada}

\author{Ruohong Li}
\affiliation{TRIUMF, 4004 Wesbrook Mall, Vancouver, BC V6T~2A3, Canada}

\author{Ryan M. L. McFadden}
\affiliation{Department of Chemistry, University of British Columbia, Vancouver, BC V6T~1Z1, Canada}
\affiliation{Stewart Blusson Quantum Matter Institute, University of British Columbia, Vancouver, BC V6T~1Z4, Canada}

\author{Gerald D. Morris}
\affiliation{TRIUMF, 4004 Wesbrook Mall, Vancouver, BC V6T~2A3, Canada}

\author{Matthew R. Pearson}
\affiliation{TRIUMF, 4004 Wesbrook Mall, Vancouver, BC V6T~2A3, Canada}

\author{Monika Stachura}
\affiliation{TRIUMF, 4004 Wesbrook Mall, Vancouver, BC V6T~2A3, Canada}

\author{John O. Ticknor}
\affiliation{Department of Chemistry, University of British Columbia, Vancouver, BC V6T~1Z1, Canada}
\affiliation{Stewart Blusson Quantum Matter Institute, University of British Columbia, Vancouver, BC V6T~1Z4, Canada}

\author{Georg Cristiani}
\affiliation{Max Planck Institute for Solid State Research, 70569 Stuttgart, Germany}

\author{Gennady Logvenov}
\affiliation{Max Planck Institute for Solid State Research, 70569 Stuttgart, Germany}

\author{Friedrike Wrobel}
\affiliation{Max Planck Institute for Solid State Research, 70569 Stuttgart, Germany}

\author{Bernhard Keimer}
\affiliation{Max Planck Institute for Solid State Research, 70569 Stuttgart, Germany}

\author{Junjie Zhang}
\altaffiliation{Current address: Materials Science and Technology Division, Oak Ridge National Laboratory, Oak Ridge, TN~37830, United States of America}
\affiliation{Materials Science Division, Argonne National Laboratory, Argonne, IL~60439, United States of America}

\author{John F. Mitchell}
\affiliation{Materials Science Division, Argonne National Laboratory, Argonne, IL~60439, United States of America}

\author{W. Andrew MacFarlane}
\email[Email: ]{wam@chem.ubc.ca}
\affiliation{Department of Chemistry, University of British Columbia, Vancouver, BC V6T~1Z1, Canada}
\affiliation{Stewart Blusson Quantum Matter Institute, University of British Columbia, Vancouver, BC V6T~1Z4, Canada}
\affiliation{TRIUMF, 4004 Wesbrook Mall, Vancouver, BC V6T~2A3, Canada}

\date{\today}

\newcommand{\latin}[1]{\emph{#1}}

\begin{abstract}
We report $\beta$-detected NMR of ion-implanted \eli\ in a single crystal and thin film of the strongly correlated metal \lno.
In both samples, spin-lattice relaxation measurements reveal two distinct local metallic environments, as is evident from $T$-linear Korringa $1/T_{1}$ below \SI{200}{\kelvin} with slopes comparable to other metals.
A small, approximately temperature independent Knight shift of \SI{\sim 74}{ppm} is observed, yielding a normalized Korringa product characteristic of substantial antiferromagnetic correlations, but, we find no evidence for a magnetic transition from \num{4} to \SI{310}{\kelvin}.
Two distinct, equally abundant \eli\ sites is inconsistent with the widely accepted rhombohedral structure of \lno, but cannot be simply explained by either of the common alternative orthorhombic or monoclinic distortions.
\end{abstract}

\maketitle

\section{Introduction \label{sec:introduction}}

On the path to understanding and controlling correlated electrons in solids, a great deal of effort has gone into studying how the Fermi liquid state can be destabilized to yield other more exotic ground states below a metal-insulator transition (MIT)\cite{Imada1998}. 
The rare-earth ($R$) perovskite nickelates \rno\ are a unique and important example that remain challenging despite intense scrutiny\cite{Torrance1992, Greenblatt1997, Catalan2008, Catalano2018}. 
Besides the MIT, further interest in \rno\ stems from their close relation to the high-$T_c$ cuprates, both as potential superconductors\cite{Anisimov1999,Chaloupka2008}, and for what the absence of nickelate superconductivity reveals about the cuprates. 
Among the \rno\ series, \lno\ (LNO) is unique in avoiding the MIT, remaining a paramagnetic metal to low temperature, making it particularly interesting and potentially useful. 
Though metallic, \lno\ is highly correlated\cite{Sreedhar1992,Stemmer2018}, with a strongly enhanced magnetic response\cite{Zhou2014} and electronic heat capacity\cite{Sreedhar1992}. 
The origin of these properties and even the persistence of the metallic state itself remain open questions\cite{Shamblin2018,Li2015}.

Recently, epitaxial strain and dimensional confinement have been used to modify the properties of \rno\cite{Catalano2018,Jak2014}, including causing an MIT in \lno\cite{Boris2011}.
Advances in high pressure O$_2$ crystal growth have also made high quality single crystals of \lno\ available for the first time\cite{Zhang2017,Guo2018}, opening the prospect for refined studies of the bulk.
Surprisingly, recent results on one crystal have cast doubt on LNO's characterization as a nonmagnetic metal, instead concluding that, in sufficiently pure stoichiometric form, it is magnetically ordered below $T_N \approx$ \SI{157}{\kelvin}\cite{Guo2018}. 
Other crystals show the propensity for slight substoichiometry and the ordering of oxygen vacancies into defect phases that may explain the observed magnetism\cite{Wang2018}.

Here we report results from $\beta$-detected NMR (\bnmr) measurements of implanted highly polarized \elip\ ions in two very different samples of LNO -- a high quality single crystal and an epitaxial thin film. 
Like muon spin rotation (\msr)\cite{Schenck1985}, \bnmr\ provides a sensitive local magnetic probe of solids, but, in contrast, due to the much longer lifetime,
it is sensitive to the metallic state in close analogy with conventional NMR\cite{Walstedt2008}.
Using a low energy beam of implanted nuclei, \bnmr\ has the additional capability that it can easily be applied to thin films\cite{2015-MacFarlane-SSNMR-68-1}.
In both samples, we find clear indication of conventional Korringa spin-lattice relaxation (SLR) below \SI{200}{\kelvin}, with no evidence of a magnetic transition from \num{4} to \SI{310}{\kelvin}. 
We find two distinct, equally abundant, metallic local environments for the implanted probe, a robust feature that, so far, defies explanation.
The quantitative similarity between the two samples strongly suggests all of these features are intrinsic.

\section{Experimental \label{sec:experiment}}

In $\beta$-NMR, highly spin-polarized $\beta$-radioactive ions are implanted into the sample, and the NMR is detected by the subsequent $\beta$-decay.
The $\beta$-decay asymmetry is proportional to the average longitudinal spin-polarization, with a proportionality constant $A_0$ that depends on the detection geometry and properties of the decay\cite{Morris2014}.
The asymmetry is measured by combining $\beta$ count rates from two opposing scintillation detectors.
All of the experiments were conducted using the $\beta$-NMR spectrometer at TRIUMF in Vancouver, Canada\cite{Morris2004,Morris2014}.
The \elip\ probe nuclei (spin $I = 2$, gyromagnetic ratio $\gamma/2\pi = \SI{6.3016}{\mega\hertz \per \tesla}$, electric quadrupole moment $Q = + \SI{32.6}{\milli\barn}$, and radioactive lifetime $\tau = \SI{1.21}{\second}$) were spin-polarized in-flight using optical pumping\cite{Levy2014} and subsequently ion-implanted into the LNO samples. 
The implantation energies were \SI{2.9}{\kilo\electronvolt} and \SI{27.9}{\kilo\electronvolt} corresponding to mean depths of \SI{12.9}{\nano\meter} and \SI{99.5}{\nano\meter} (details in Appendix \ref{app:Beam-energy}).

In an applied magnetic field $B_0 =$ \SI{6.55}{\tesla} provided by a high homogeneity superconducting solenoid in persistence mode, two types of measurements were performed: relaxation and resonance.
With a pulsed \elip\ beam, SLR data were collected by monitoring the depolarization during and after the \num{4} second long pulse.
During the pulse, the polarization approaches a dynamic steady-state value, while afterwards, it relaxes to \num{\approx 0}. 
Since the probe nucleus is polarized prior to implantation, unlike conventional NMR, no radio frequency (RF) field is required to measure SLR. 
Resonance measurements used a continuous beam of $^8$Li$^+$ with a transverse RF field $H_1$ stepped slowly in frequency through the $^8$Li Larmor frequency $\nu_0 = \gamma B_0 /(2\pi) \approx$ \SI{41.27}{\mega\hertz}. 
On resonance, the $^8$Li spin precesses rapidly due to the RF field, resulting in a loss of the time-averaged asymmetry. 
The resonance frequency was calibrated against a single crystal of MgO at \SI{300}{\kelvin}\cite{MacFarlane-2014-MgO}.

The film sample was deposited on a single crystal [LaAlO$_3$]$_{0.3}$[Sr$_2$AlTaO$_6$]$_{0.7}$ (LSAT) substrate by pulsed laser deposition (PLD) and annealed in an O$_2$ rich environment as described in Ref.\ \onlinecite{Boris2011}. Its thickness was determined by X-ray reflectivity to be \SI{38+-1}{\nano\meter}.
The LSAT substrate is lattice matched to LNO, minimizing epitaxial strain. 
The LNO crystal was grown as detailed in Ref.\ \onlinecite{Zhang2017}.
A \SI{1.5}{\milli\meter} thick slice perpendicular to the pseudocubic [100] direction was cut from a cylindrical boule \SI{\sim 5}{\milli\meter} in diameter.
The surface was prepared by polishing with Al$_2$O$_3$ suspensions until a mirror-like surface was obtained.
The samples were affixed to polished sapphire plates attached to a He cold-finger cryostat.

Although the crystals have a rather low average O deficiency (determined by TGA to be \num{\sim 0.015}/formula unit), it is possible that this could influence the measured properties. 
Indeed, large O vacancy concentrations can be purposely generated in \lno\ with the appearance of supercell defect phases that order magnetically, with composition LaNiO$_{2.75}$ and LaNiO$_{2.5}$\cite{Wang2018}.
If the O vacancy phases are uniformly distributed, they would amount to <5~mol~\%; however, in the case of nonuniform distribution, the concentration of ordered superlative phases would be sample-dependent and could exceed this upper bound. 

\section{Results and Analysis \label{sec:results}}

\subsection{Spin-Lattice Relaxation}

\begin{figure}[ht]
\includegraphics[width = 1.0\columnwidth]{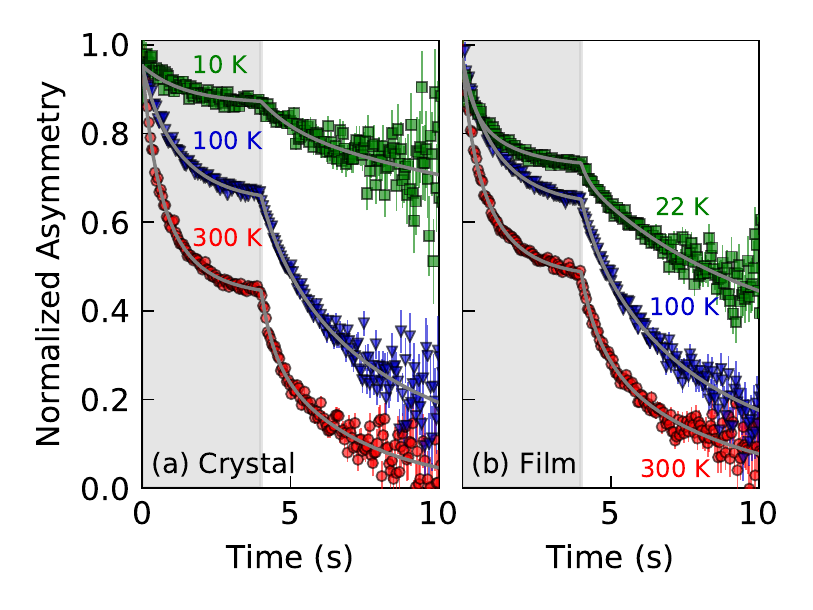}
\caption{Examples of SLR data for \elip\ in LaNiO$_3$ with $B_{0}$ = \SI{6.55}{\tesla} with fits (lines) using the triexponential of Eq.\ \eqref{eq:1} convoluted with the \SI{4}{\second} beam pulse indicated by the shaded area.
The data are normalized by their apparent $t = 0$ asymmetry, see Appendix \ref{App-2}.}
\label{fig:SLR-fit-ex}
\end{figure}

Representative SLR data are shown in Figure \ref{fig:SLR-fit-ex}. 
At low temperatures in both the crystal and film, the relaxation is slowest, with the rate increasing monotonically with temperature. 
We identify a small amplitude fast relaxing component at early times as a background signal due to \eli\ stopping outside the sample. 
It is easily distinguished from the sample signal with a nearly temperature independent rate faster than the sample at all temperatures.
The remaining signal from LNO is also not comprised of a single relaxing component, but has two distinct SLR contributions with different rates. 
With this in mind, we require a model relaxation function $R(t)$, the analog of the magnetization recovery curve in NMR, to fit the data. 
The simplest form providing a good fit is a triexponential, which encapsulates the biexponential signal from LNO, as well as the background.
Specifically, at time $t$ after an \eli\ arriving at time $t'$,
\begin{equation} \label{eq:1}
\begin{split}
   R(t,t') &= f_{S}[ (1-f_{f})e^{-\lambda_{s}(t-t')} + f_fe^{-\lambda_{f}(t-t')}] \\ &+ (1-f_S)e^{-\lambda_{b}(t-t')},
\end{split}
\end{equation}
where $\lambda_{i} \equiv 1/T_{1}^{i}$ ($i = s, f, b$) are the SLR rates, $f_{f} \in [0,1]$ is the fast relaxing fraction, $f_{S}$ is the  fraction of \eli\ in the sample, and the third term is the background.
In order to fit the data, a global procedure was used wherein all spectra for each sample (i.e., at every temperature) were fit simultaneously using custom C++ code and the MINUIT minimization routines\cite{minuit} provided by the ROOT framework\cite{root}.
Best fits were obtained using temperature independent fractions: $f_f=$\num{0.50+-0.03} for both samples and $f_S =$ \num{0.85+-0.06} for the crystal and \num{0.80+-0.06} for the film (see Appendix \ref{App-2}), leaving $\lambda_i$ as the only temperature dependent parameters. 
The fit quality is good in each case: $\chi^{2}$ = \num{0.94} (crystal) and \num{1.05} (film).

\begin{figure}[ht]
\includegraphics[width = 1.0\columnwidth]{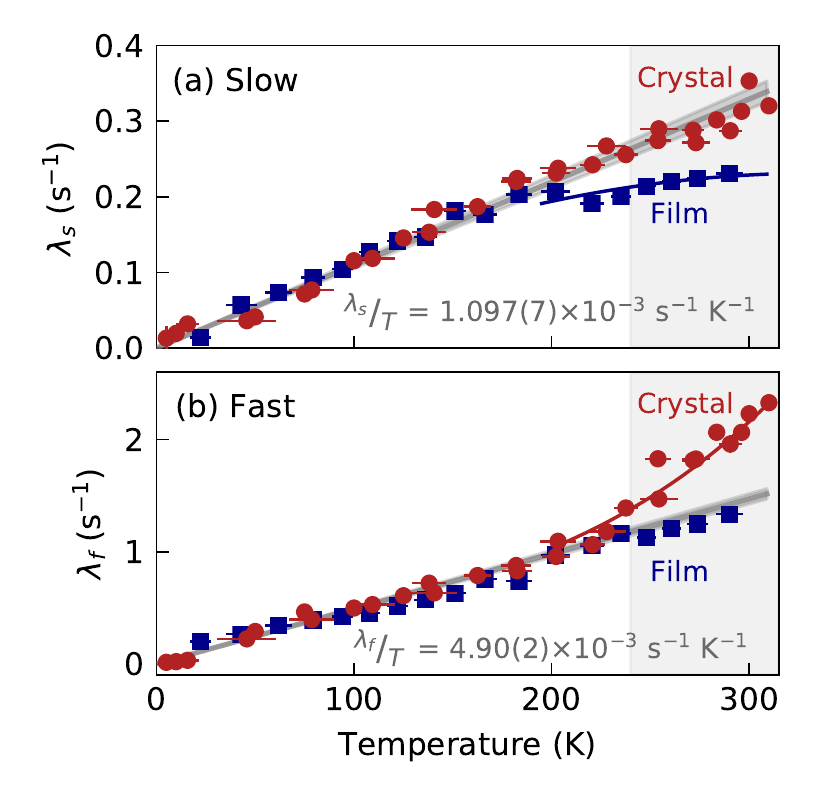}
\caption{The slow (a) and fast (b) SLR rates as a function of temperature for $^8$Li in the LaNiO$_3$ crystal (red circles) and \SI{38}{\nano\meter} film (blue squares).
Both rates exhibit a linear Korringa behavior below \SI{200}{\kelvin} that agrees {\it quantitatively} between the two samples.
The thick grey lines are the best fit slopes, with the values shown.
The shading shows the region bounded by a $1\sigma$ change to the slope.
The vertical shaded region marks the onset of the high temperature decrease in the single crystal $\chi_0(T)$ \cite{Zhang2017} (see text).}
\label{fig:SLR-Tdep}
\end{figure}

The slow and fast SLR rates extracted from the above analysis are shown as a function of temperature in Fig.\ \ref{fig:SLR-Tdep}(a) and (b). 
Consistent with the qualitative features in Figure \ref{fig:SLR-fit-ex}, the rates are slowest at the lowest temperature and increase linearly up to \SI{\sim 200}{\kelvin}, above which there are sample dependent nonlinearities.
Significantly, the fast and slow rates are quantitatively consistent between the two samples indicating the behavior is \textit{intrinsic}.
Over the linear range ($T<$ \SI{200}{\kelvin}), we fit $\lambda_i(T)$ to a line to obtain the slopes in Fig.\ \ref{fig:SLR-Tdep}.
The two components are present in equal amplitudes in both samples with rates that differ by a factor of \num{\sim 4} on average. Their origin is discussed in Section \ref{discussion}.

\subsection{Resonances}

\begin{figure}[ht]
\includegraphics[width = 1.0\columnwidth]{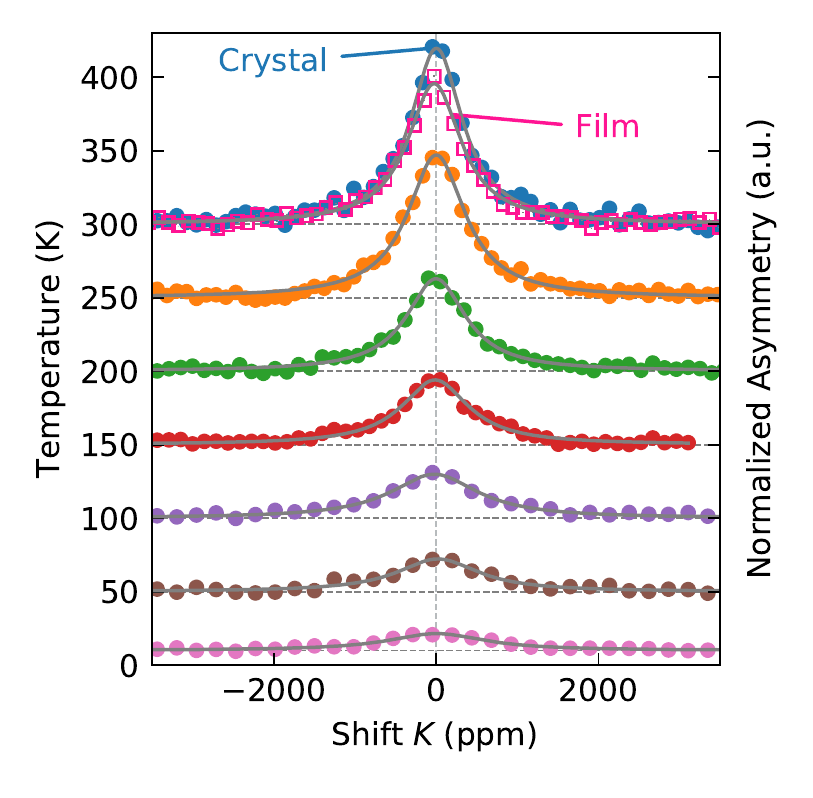}
\caption{Temperature dependence of the \eli\ resonance in the LaNiO$_3$ crystal at \SI{6.55}{\tesla}.
At low temperature, the resonance is broadest and weakest, with the amplitude growing monotonically and the width narrowing by a factor of \num{\sim 2} as the temperature is raised. 
A small, nearly temperature independent positive shift relative to MgO is evident. 
The spectra are normalized by the off-resonance baseline asymmetry and offset to clearly display the temperature dependence.
The \eli\ resonance in the LNO film at \SI{300}{\kelvin} is also shown (open squares); it is remarkably similar to the crystal both in shift $K$ and width. }
\label{fig:Res-fit-ex}
\end{figure}

\begin{figure}[hb]
\includegraphics[width = \columnwidth]{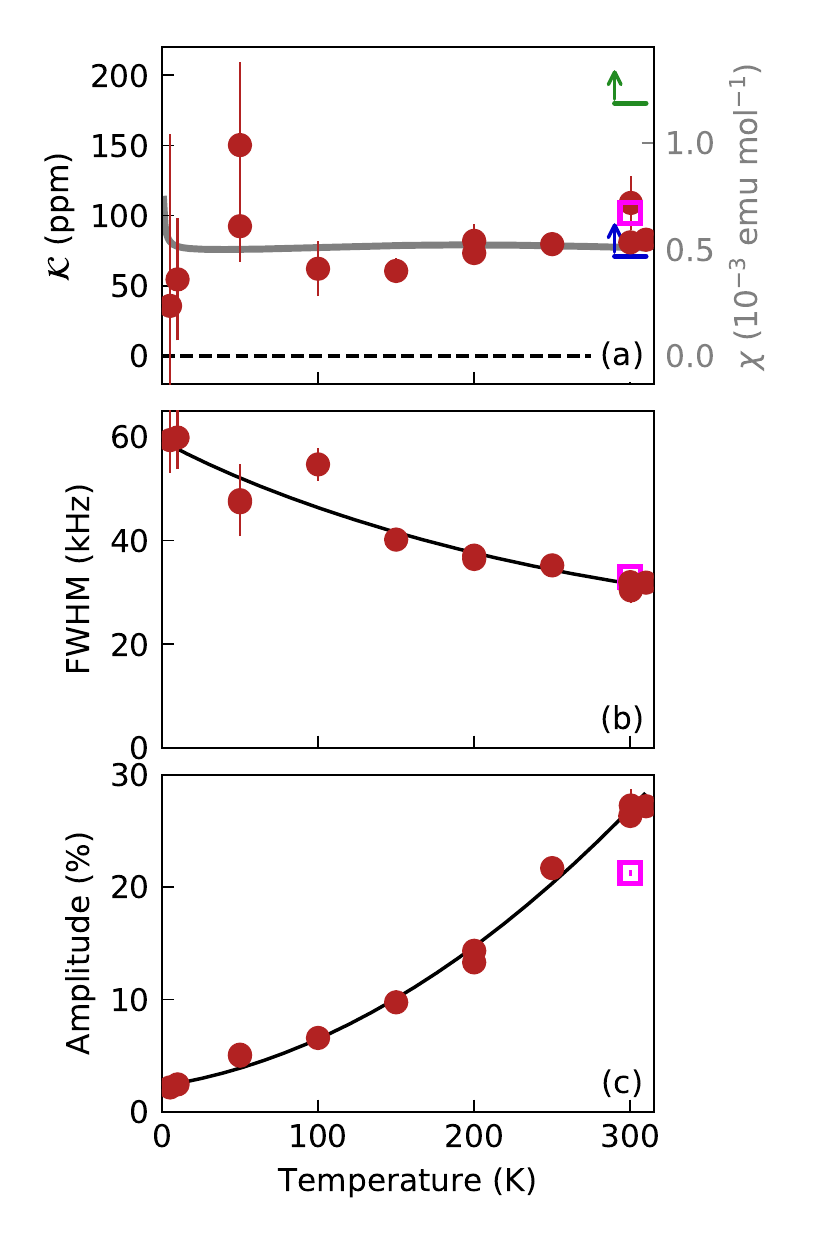}
\caption{Temperature dependence of the \elip\ resonance in the LaNiO$_3$ crystal:
(a) Knight shift $\mathcal{K}$, (b) full-width half maximum (FWHM), and (c) amplitude.
The amplitude is normalized by the off-resonance baseline asymmetry. 
In (a), the solid grey curve and right scale show the magnetic susceptibility of the crystal\cite{Zhang2017}.
Horizontal lines show estimates of the Knight shift for the slow (blue) and fast (green) relaxing components at \SI{300}{\kelvin}.
The vertical arrows emphasize that the values are lower limits (see text).
The black solid curves in panels (b) and (c) are guides. 
The values for the \SI{38}{\nano\meter} film at \SI{300}{\kelvin} are also shown (open squares).}
\label{fig:Res-Tdep}
\end{figure}

The resonances (primarily in the crystal) are shown in Fig.\ \ref{fig:Res-fit-ex}. 
In contrast to the multiexponential SLR, the spectrum consists of a single broad line which is narrowest and most intense at the highest temperature. 
The spectra were fit to a single Lorentzian, and the resulting parameters are shown in Fig.\ \ref{fig:Res-Tdep}.
As the temperature is lowered, the peak broadens by a factor of $\sim$\num{2},
while its amplitude decreases by an order of magnitude.
Remarkably, the resonance in the film at \SI{300}{\kelvin} is comparable to the crystal.
Aside from the breadth, the resonance is also slightly positively shifted from the calibration in MgO. 
We quantify its relative shift by
\begin{equation} \label{eq:2}
K \mbox{~[ppm]} = \left( \frac{\nu_{\mathrm{LNO}}-\nu_{\mathrm{MgO}}}{\nu_{\mathrm{MgO}}} \right) \times 10^{6}.
\end{equation}
From which we extract the Knight shift $\mathcal{K}$, due to the conduction band spin susceptibility, after accounting for demagnetization (see Appendix \ref{App-3}).  
The resulting $\mathcal{K} \sim$ \SI{74}{ppm}, independent of temperature, is typical of \eli\ in metals but is surprisingly small considering the rather large susceptibility. 
We discuss the shift further below.

\section{Discussion \label{discussion}} 

Like other atomic solutes, the implanted \eli\ nucleus generally has some hyperfine coupling to the conduction band of the host, resulting in a phenomenology very similar to the conventional NMR of metals\cite{Walstedt2008}. 
The predominant SLR mechanism is usually the Korringa process of spin exchange scattering with the conduction electrons, leading to a characteristic linear dependence of $1/T_1$ on temperature\cite{2015-MacFarlane-SSNMR-68-1}. 
One distinction from conventional NMR is that the lattice site of the implanted ion is not known {\it a priori} and is not independently accessible by X-ray diffraction. 
The site is important because it determines the sensitivity of $T_1$ to spin fluctuations at finite wavevector, such as when there is some tendency towards antiferromagnetism. 
The relaxation rate is related to the generalized frequency- and wavevector- dependent magnetic susceptibility $\chi = \chi' - i \chi''$ via the Moriya expression\cite{Moriya1963},
\begin{equation} \label{Moriya}
\frac{1}{T_1} = \frac{4 k_B T}{\hbar} \sum_{\vec{q}} \frac{|A(\vec{q})|^2}{(\gamma_e \hbar)^2} \frac{1}{\hbar \omega_0} \chi''(\vec{q},\omega_0),
\end{equation}
where the hyperfine form factor $A(\vec{q})$, the spatial Fourier transform of the local hyperfine coupling, acts as a filtering function on the magnetic fluctuations at wavevector $\vec{q}$ and the NMR frequency $\omega_0$. $\gamma_e$ is the electron gyromagnetic ratio.
The resulting site dependence of $1/T_1$ is illustrated by the well known case of NMR in the cuprates\cite{Walstedt2008}.

Another important difference of \eli\ \bnmr\ from, e.g.\ transition metal NMR, is that, as a light isotope, the hyperfine coupling is quite weak, similar to the implanted muon\cite{Schenck1985}, dissolved hydrogen\cite{Barnes1997} or stable $^7$Li\cite{MacFarlane2000}, leading to
small Knight shifts on the order of \SI{100}{ppm} and correspondingly long $T_1$.
The small \eli\ shifts make the demagnetization correction particularly important\cite{Xu2008}. 
Unlike the muon\cite{Blundell2001}, the Korringa rate is usually in the accessible window for \eli: $0.01 \tau < T_1 < 100 \tau$, making it a useful probe of metals, including correlated oxides such as Sr$_2$RuO$_4$\cite{Cortie2015}.

In the present data, the linearity of $1/T_1$ for both fast and slow components below \SI{200}{\kelvin} (Fig.\ \ref{fig:SLR-Tdep}) is clear microscopic evidence of a conventional metallic state, consistent with $^{139}$La NMR of bulk powder\cite{Sakai2002}. 
Recent powder neutron diffraction suggest that LNO is electronically inhomogeneous, possessing insulating pockets below \SI{200}{\kelvin} \cite{Li2015}. 
In contrast, we find no evidence for an insulating volume fraction.

\bnmr\ reports the average behavior over the sampled volume given by the implantation profile (Appendix \ref{app:Beam-energy}) and the millimetric beamspot. 
An insulating LNO phase would produce another SLR component with amplitude proportional to its volume fraction and  a relaxation distinct in both magnitude and temperature dependence. 
For putative insulating regions, we expect strong local moment magnetism and low temperature magnetic order. 
In such regions, the relaxation would be fast and increasing with reduced temperature, peaking at the magnetic freezing, where it might even be so large as to ``wipe out'' the signal. 
Based on this, we rule out an insulating magnetic fraction greater than an estimated detection limit of $\sim$\SI{5}{\%}.

The \eli\ resonance spectrum characterizes the static, time-average magnetic properties.
Surprisingly, we find a single broad line showing none of the quadrupolar fine structure observed in related insulating perovskites\cite{STO2003,LSAT-2017,LAO-2017}. 
A reduction of the electric field gradient (EFG) compared to isovalent LaAlO$_3$ (LAO)\cite{LAO-2017} is reasonable in the negative charge transfer picture, as the extra $+1$ charge of Al$^{3+}$ (vs.\ Ni$^{2+}$) is spread over the oxide ligands. 
Metallic screening will further reduce the EFG, evidently resulting in an unresolved splitting of the NMR on the order of the linewidth or less.

The line is, in fact, very broad compared to both other metals and insulating perovskites\cite{STO2003,KarnerLAO}. 
The temperature dependence of the width [Fig.\ \ref{fig:Res-Tdep}(b)] suggests a combination of a substantial temperature independent term and one that increases as temperature is decreased. 
Qualitatively, the former is consistent with quadrupolar broadening due to structural disorder and the latter to magnetic broadening from dilute magnetic defects. 
However, significantly, the widths in the two samples are comparable at \SI{300}{\kelvin}, the only temperature where we have a resonance in the film. 
One would not expect extrinsic disorder to be very similar in these vastly different samples, so this agreement is surprising and also points to an intrinsic origin for the width. 

To understand these features, we must consider the structural details of perovskites. 
The nominal rhombohedral structure involves a rotation of the NiO$_6$ octahedra about any of the equivalent $\langle 111 \rangle$ directions of the cubic phase.
As a result, the crystal is microtwinned - an intrinsic inhomogeneity that may contribute to the quadrupolar broadening.
However, similar microtwinning is also a feature of rhombohedral LAO, where we find large well-resolved quadrupole splitting and substantially smaller linewidth\cite{KarnerLAO}, as well as tetragonal SrTiO$_3$\cite{STO2003}.
This suggests twinning alone cannot be responsible for broadening.
We return to this point below.

There is also no evidence in the resonance spectra for two environments corresponding to the two relaxing components, but this is not surprising on quantitative grounds, since the observed shift is so small [Fig.\ \ref{fig:Res-Tdep}(a)].
The relaxation rate is determined by the square of the hyperfine coupling [Eq.\ \eqref{Moriya}], while the shift is only linear, so the factor of \num{4} between the rates implies only a factor of \num{2} in the shifts, yielding a magnetic splitting on the order of \SI{50}{ppm} (\SI{2}{\kilo\hertz} at this $B_0$, significantly less than the linewidth). 
Moreover, though the relaxing components are practically equal in their initial (i.e., $t=0$) asymmetry, the resonance amplitude is determined instead by the time average asymmetry\cite{Hossain2009}, which, for the fast relaxing component, is suppressed to at most \SI{27}{\%} of the total. 
Finally, if there is some magnetic broadening, the larger coupling to the fast relaxing component would likely make its resonance broader and even smaller in relative amplitude.

Based on these considerations, we conclude that the \eli\ \bnmr\ is an unresolved composite of two lines, originating from two distinct local environments, whose spectrum is heavily weighted towards the slow component.
The resonance shift (\SI{\sim 74}{ppm}) is then a weighted average of the two.
To illustrate this, we decompose the spectrum at \SI{300}{\kelvin}, assuming equal linewidths and using the relative amplitudes from $\lambda$ as weights\cite{Hossain2009}, to estimate the Knight shifts of the two components. 
These are shown in Fig. \ref{fig:Res-Tdep}(a), as the horizontal lines for the slow (blue) and fast (green) component where the vertical arrows emphasize that they are lower limits, since the fast relaxing resonance may be wider (and hence smaller). 

From this analysis, we estimate the Knight shift $\mathcal{K} \approx$ \SI{71 \pm 10}{ppm} for the major (slow) component at \SI{300}{\kelvin}. 
Combining this with the Korringa slope [Fig.\ \ref{fig:SLR-Tdep}(a)], we form the normalized Korringa product,\cite{Walstedt2008}
\begin{equation}
\mathscr{K} = \frac{T_{1}T\mathcal{K}^{2}}{S} = \num{0.40 \pm 0.10},
\end{equation}
where for \eli\, $S = $ \SI{1.20E-5}{\second\kelvin}.
For an uncorrelated free electron metal, $\mathscr{K} \approx 1$. 
Our $\mathscr{K}$ is significantly less, indicating substantial antiferromagnetic correlations\cite{Moriya1963,Ueda1975}.
This is opposite to a recent Stoner enhancement interpretation of the susceptibility of powder\cite{Zhou2014}, but it is consistent with the occurrence of antiferromagnetism in the insulating nickelates\cite{GarciaMunoz1995}, including thin layers of LNO\cite{Frano2013}. 
In contrast, $^{139}$La NMR finds $\mathscr{K} \approx 1.5$ (i.e.\ a relaxation rate {\it slower} than expected from $\mathscr{K} = 1$) suggesting instead ferromagnetic correlations\cite{Sakai2002}.

We now consider potential sources for this substantial discrepancy. 
One possibility is that, due to the different lattice sites, the different form factors in Eq.\ \eqref{Moriya} blind the La to an important wavevector that implanted \eli\ sees, reducing its $T_1$ and $\mathscr{K}$.
The most obvious candidate is the reported AF ordering vector in LNO/LAO superlattices,
$\vec{q}_{\mathrm{AF}} = (1/4,1/4,1/4)$\cite{Frano2013}. However, for the La site $A(\vec{q}_{\mathrm{AF}})$ is nonzero, ruling out at least the most obvious explanation along these lines.
A second possibility is that (unlike Li), La may have a substantial orbital shift which cannot easily be separated from the temperature independent Knight shift. Using the full $^{139}$La shift may thus result in a significantly overestimated $\mathscr{K}$. 

While it is widely accepted that LNO remains a metal to low temperature, the occurrence of static antiferromagnetism (AF) was recently suggested from new data in a single crystal at $T_N \approx$ \SI{157}{\kelvin}.\cite{Guo2018} 
In contrast, we have no evidence of magnetic ordering at this temperature, either from the resonance line or $1/T_1$, in agreement
with recent data suggesting that the AF is due to an oxygen deficient phase.\cite{Wang2018}

We turn now to the sample dependent deviation from the Korringa dependence of $1/T_1$ above \SI{200}{\kelvin}.
This bifurcation coincides with a small change in the Ni-O-Ni angle from structural studies on powders\cite{Li2015}.
Above about \SI{300}{\kelvin}, $\chi(T)$ for LNO crystals is well described by the phenomenological Curie-Weiss dependence, $\chi \propto (T + \theta)^{-1}$ with a large $\theta \approx$ \SI{2000}{\kelvin}, indicating a substantial departure from a simple T independent Pauli susceptibility.
It is interesting that the Korringa dependence in Fig. \ref{fig:SLR-Tdep} (the NMR hallmark of a metal) appears to break down close to the onset of this high $T$ decrease in $\chi$.
From analysis of the RF conductivity\cite{Shamblin2018}, it was concluded that the carrier density decreases substantially above \SI{200}{\kelvin} which would simultaneously diminish both Korringa slopes.
The sample dependence is clearly more complex (Fig. \ref{fig:SLR-Tdep}).
It is also not as simple as an additional relaxation, due to differing impurity content, for example.
Rather, it has the opposite sense for the fast and slow components. 
This suggests the magnetic response depends on subtle details of the temperature evolution of the lattice which are in turn influenced by the epitaxial relation to the substrate.

\begin{figure}[ht]
\includegraphics[width = 0.9\columnwidth]{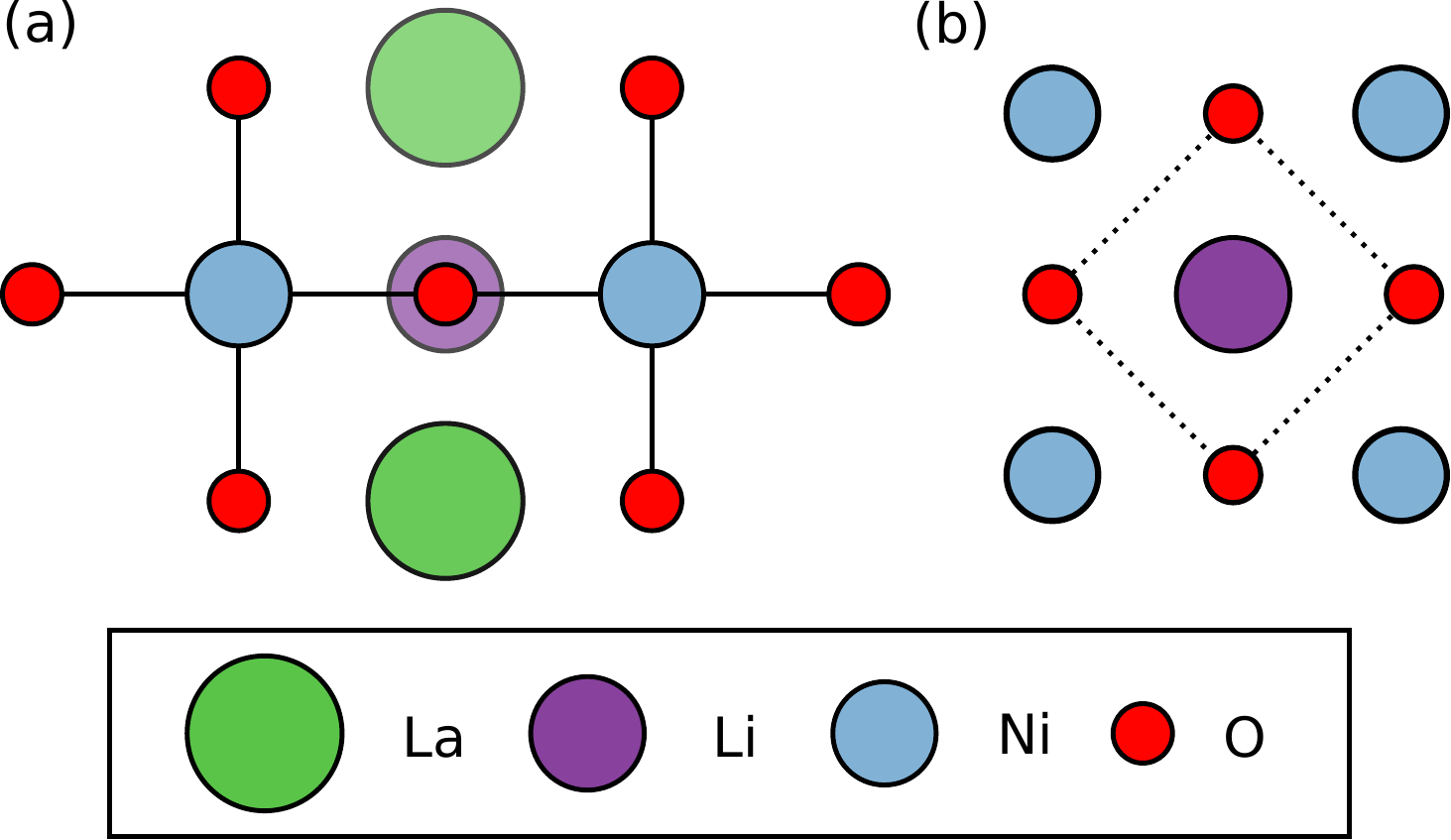}
\caption{Schematic of the interstitial \elip\ site $P$ in perovskite oxides. 
For simplicity, the illustrated structure is cubic; however, the site retains the same coordination when the symmetry is lowered. 
(a) Side view illustrating that the Li site is halfway between two neighbouring La. 
(b) Top view of the NiO$_2$ plane showing that the Li site is coordinated with 4 oxygen anions. }
\label{fig:Cubic-view}
\end{figure}

The above features are consistent with a correlated metal without static magnetism. 
However, an essential feature of our data is the two component character of the relaxation indicating two metallic local environments for the \eli. 
Based on the \bnmr\ of \eli\ in simple metals\cite{Morris2004,Parolin2007,Zalman2007,Parolin2008,Parolin2009,Hossain2009,Ofer-Pt-2012}, one possibility is simply that there are two inequivalent sites for the implanted \eli\ within the unit cell. 
In elemental Ag and Au\cite{Morris2004,Parolin2008,Hossain2009}, for example, there are two cubic \eli\ sites that differ in their hyperfine coupling to the conduction band, and we find a $T$ dependent site change transition around \SI{180}{\kelvin}\cite{Parolin2008}. 
In other perovskite oxides\cite{STO2003,LSAT-2017,LAO-2017}, we find a single interstitial \elip\ site characterized by a large quadrupole splitting\cite{STO2003,LSAT-2017,KarnerLAO}, which we shall call the $P$ site, at the $3c$ Wyckoff position in the cubic ($Pm\bar{3}m$) phase.
As illustrated in Fig. \ref{fig:Cubic-view}, $P$ is midway between two adjacent $A$-site (La$^{3+}$) ions at the centre of a square Ni-O plaquette. 
In contrast to Au and Ag\cite{Morris2004,Parolin2008,Hossain2009}, the two components in LNO are equal in amplitude at all temperatures, and we find no hint of a site change.
Equal population of two distinct lattice sites over such a wide range of $T$, while possible, seems unlikely.

With no obvious second site, we now consider how site $P$ evolves as the crystal symmetry is lowered from the ideal cubic perovskite structure.
In the rhombohedral phase (the nominal LNO structure), adjacent NiO$_6$ octahedra rotate with an equal angle $\sim 9^\circ$ and alternating sign\cite{Gou2011}.
Despite this, there remains a single $P$ site, consistent with the \eli\ spectrum in rhombohedral LAO.\cite{KarnerLAO}
In the orthorhombic ($Pnma$) structure, the rotation pattern of the NiO$_6$ octahedra differs ($a^-a^-c^+$ instead of $a^-a^-a^-$, in Glazer notation\cite{Glazer1972}), resulting in three distinct $P$-derived Li sites and two distinct oxygen sites in the enlarged unit cell.
With a further lowering of symmetry to monoclinic with the rocksalt alternating superlattice of long and short bond NiO$_6$ octahedra,
i.e.,\ the bond disproportionated structure of insulating \rno,
$P$ further diverges into 4 distinct sites (while there are two Ni and three O sites).
Note that in all these structures there is a single La ($A$) site.
Since these structures are all pseudocubic, the $P$ derived sites are all very similar, with comparable
interstitial space to accommodate the implanted ion but slight differences in distances to the near neighbours and in Li-O-Ni angles that in the cubic phase are exactly $90^\circ$. 
Supercell density functional theory (DFT) calculations (Appendix \ref{App-4}) confirm that energetically they are all very similar, so that one would expect a randomly implanted \elip\ to occupy them with equal probability. 
While a distribution of similar (but distinct) sites in a lower symmetry structure constitutes an intrinsic microscopic source of inhomogeneity that could account for the resonance width, among these three structures, we have not found an obvious explanation for two equally abundant \eli\ environments (detailed in Appendix \ref{App-5}).
Based on this, we suggest that {\it the precise low temperature structure of LNO may be none of these commonly considered possibilities}.

Interestingly, our DFT calculations predict that the ground state symmetry of LNO is rhombohedral $R\bar{3}$, where the $a^-a^-a^-$ rotation pattern of the NiO$_6$ octahedra coexists with the breathing distortion. 
A similar observation was made earlier in Ref.~\onlinecite{Subedi2018}.
In this structure, there are two $P$ derived sites with energies as close as \SI{20}{\milli\electronvolt} when occupied by \lip.
The distortions giving rise to these structural variants are still subtle and not easily distinguished even with high resolution diffraction\cite{Li2015,Shamblin2018}.
The two component character of the \eli\ relaxation demonstrates that a local probe may provide important structural insight, particularly when there is substantial microscopic inhomogeneity as evident in the resonance width.
It will be important to carry out structural refinement of the diffraction data based on the $R\bar{3}$ structure.
In addition, other local probes (such as  $^{17}$O or the very difficult $^{61}$Ni\cite{vdk2010}) would provide further insight into the local structure and properties of \lno. 

\section{Conclusion \label{sec:conclusion}}

Using ion-implanted \eli\ \bnmr, we studied the local electronic properties of \lno. 
SLR measurements revealed two components with linearly temperature dependent $1/T_1$ below \SI{200}{\kelvin}, consistent with a Korringa mechanism, providing strong microscopic evidence of a conventional metallic state.
The \eli\ resonance spectrum comprised a single broad line, implying considerable static inhomogeneity at all temperatures.
We find no evidence for either an insulating volume fraction or an antiferromagnetic ordering transition.
However, the normalized Korringa product $\mathscr{K} = \num{0.40 +- 0.10}$ indicates substantial AF correlations.
The two component SLR implies two \eli\ environments, indicating the local crystal symmetry is inconsistent with the three commonly considered possibilities: the rhombohedral $R\bar{3}c$, the orthorhombic $Pnma$ and the monoclinic $P2_{1}/n$. 
Based on DFT calculations, we postulate that the local structure of LNO is $R\bar{3}$ where the rhombohedral unit cell is expanded by a breathing distortion of the NiO$_6$ octahedra.  

\begin{acknowledgments}

We thank R. Abasalti, D.J. Arseneau, S. Daviel, B. Hitti and D. Vyas for technical assistance, and E. Benckiser, J. Chakhalian, Z. Salman and G. Sawatzky for helpful discussions.
This work was supported by NSERC Canada.  

\end{acknowledgments}

\appendix

\section{Beam Energy and Implantation Depth \label{app:Beam-energy}}

\begin{figure}[hb]
\includegraphics[width = 1.0\columnwidth]{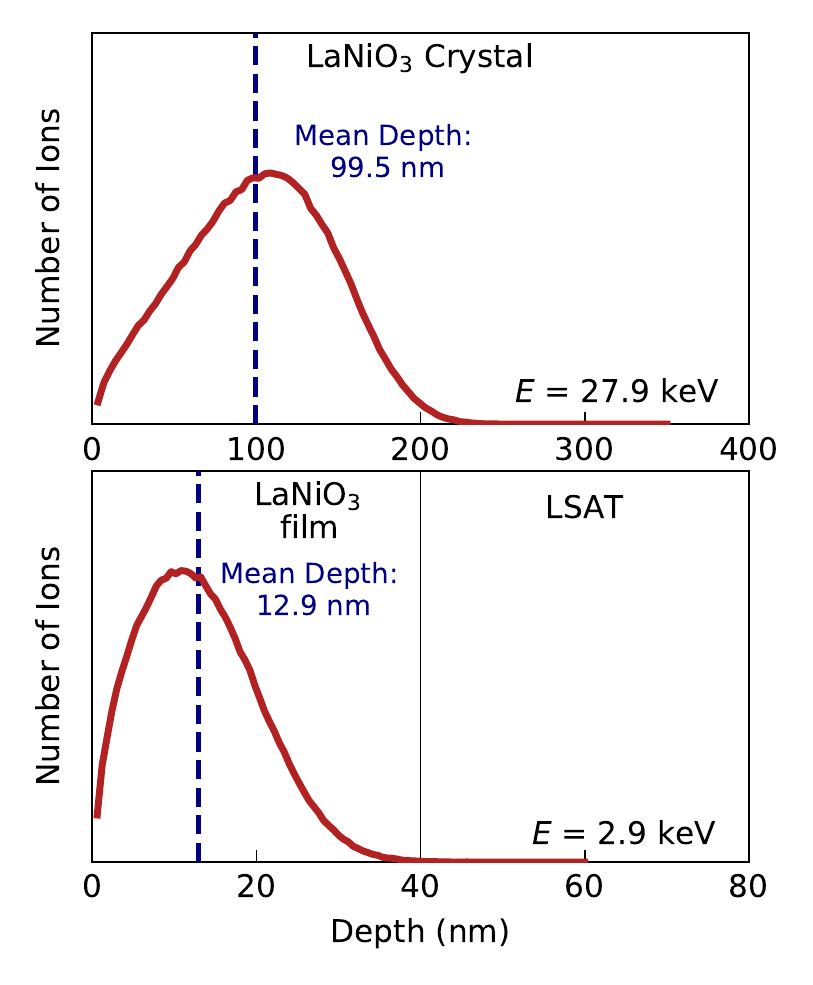}
\caption{Stopping distribution of \eli\ in a \lno\ crystal (top), and a \SI{38}{\nano\meter} film deposited on LSAT (bottom). 
Obtained using SRIM\cite{Ziegler2010} calculations for implantation energies $E$ of \SI{27.9}{\kilo\electronvolt} and \SI{2.9}{\kilo\electronvolt}.
Each simulation used $10^{5}$ Li ions.}
\label{fig:SRIM-profiles}
\end{figure}

The energy of the \elip\ beam determines the mean implantation depth of the probe into the sample. 
The \bnmr\ spectrometer is located on an isolated high voltage platform and the ion beam can be decelerated (i.e., the desired implantation depth can be chosen) by varying the bias\cite{Morris2004,Morris2014}. 
We modeled the implantation using the SRIM Monte Carlo simulation package\cite{Ziegler2010}, and results for the beam energies used here are shown in Figure \ref{fig:SRIM-profiles}.
For the \SI{38}{\nano\meter} film, the implantation energy was chosen to be \SI{2.9}{\kilo\electronvolt} (mean depth \SI{13}{\nano\meter}) to minimize the amount of \eli\ in the LSAT substrate. 
In contrast, no such concerns were warranted for the crystal and the full beam energy \SI{27.9}{\kilo\electronvolt} (mean depth \SI{100}{\nano\meter}) was used.

\section{Decomposition of the Spin-Lattice Relaxation \label{App-2}}

\begin{figure}[ht]
\includegraphics[width = 1.0\columnwidth]{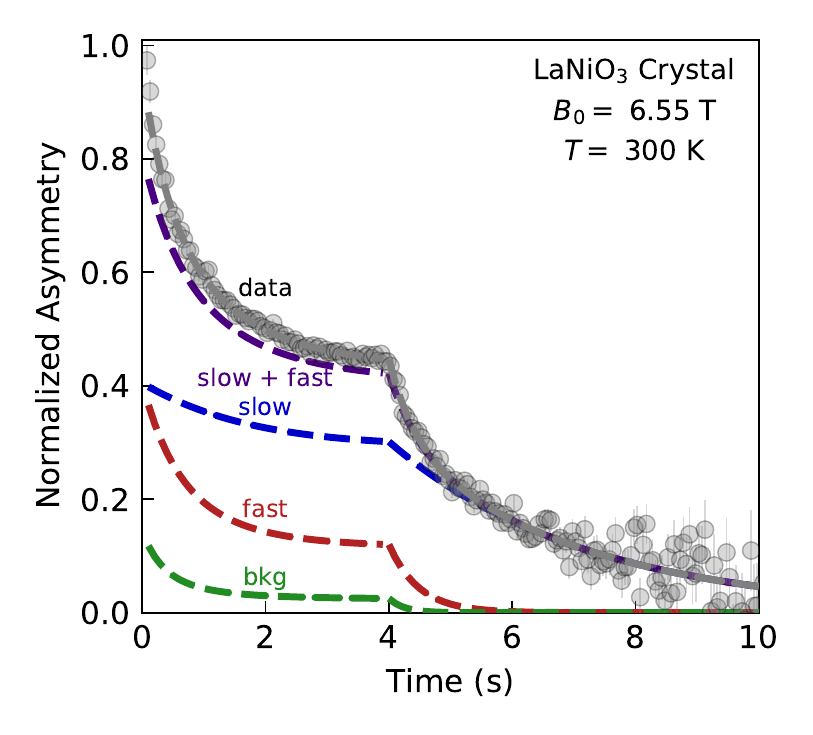}
\caption{An example decomposition of the SLR fit to Eq.\ \eqref{eq:1} for the crystal at \SI{300}{\kelvin} and \SI{6.55}{\tesla}, showing the individual components: background (green), fast (red), and slow (blue). 
Also shown: the sum of slow and fast (purple) and the total fit (grey).
}
\label{fig:bkg-demo}
\end{figure}

Here we present a more detailed view of the analysis of the SLR data.
The phenomenological fitting function based on Eq.\ \eqref{eq:1} consists of three exponentials with independent rates whose contributions are illustrated in Figure \ref{fig:bkg-demo} at \SI{300}{\kelvin}, where the relaxation is fastest.
The background component has the largest relaxation rate but by far the smallest amplitude. 
In contrast, the slow and fast sample components have equal amplitudes, but rates that differ by a factor of $\sim 4$.
Note that though small, the background signal is essential for a good fit. 
It is comparable in both amplitude and rate to similar signals that are clearly evident in samples with much slower relaxation, such as nonmagnetic insulators, e.g.\ Fig.\ 5 in Ref. \onlinecite{LSAT-2017}.
Table \ref{tab:fit-par-SLR} lists the (temperature independent) values of the full asymmetry $A_{0}$, $f_{S}$, and $f_{f}$ for the two samples, with statistical errors from the global fits. 

\begin{table}[]
\centering
\caption{The total experimental asymmetry $A_0$ and the fractions in Eq.\ \eqref{eq:1} from the global fit analysis of the SLR data in LNO.}
\begin{tabular}{ c  c  c  c }
\hline
 Sample & $A_{0}$ & $f_{S}$ & $f_{f}$ \\ 
 \hline 
 Crystal & \num{0.085 \pm 0.003}  & \num{0.85\pm 0.06} & \num{0.5 \pm 0.02} \\ 
 Film & \num{0.090 \pm 0.003}  & \num{0.80 \pm 0.06} & \num{0.5 \pm 0.03} \\ 
 \hline
\end{tabular}
\label{tab:fit-par-SLR}
\end{table}

\section{Demagnetization Correction \label{App-3}}

\begin{figure}[ht]
\includegraphics[width = 1.0\columnwidth]{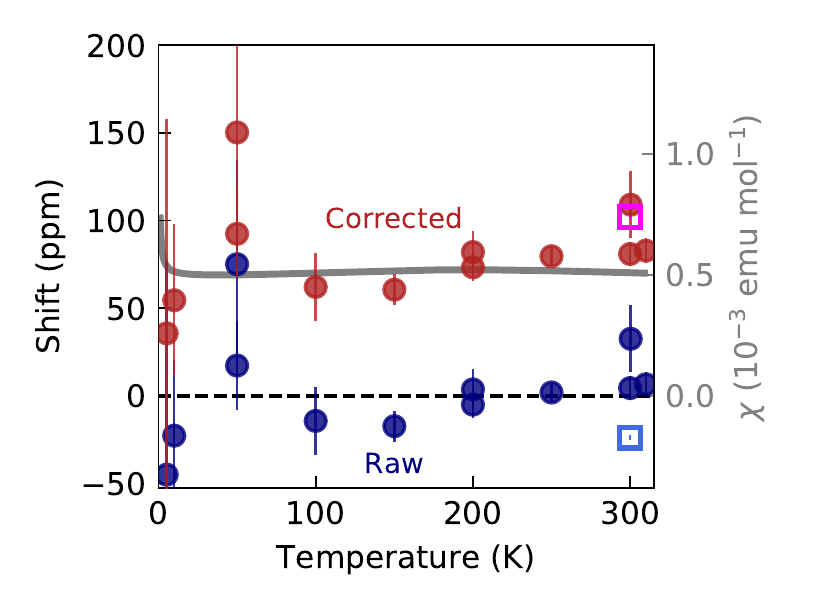}
\caption{Comparison of the raw and (demagnetization) corrected shift of the \eli\ NMR in the LaNiO$_3$ crystal. 
The raw shift is obtained by using Eq. (\ref{eq:2}), while the corrected shift is obtained by applying the demagnetization correction in Eq. (\ref{eq:demag-corr}).
The results at \SI{300}{\kelvin} for the film are shown as open squares. 
Overlaid on the corrected shifts is the molar susceptibility of the crystal\cite{Zhang2017}, right scale.}
\label{fig:raw-shift}
\end{figure}

The raw shift with respect to MgO is proportional to the static average internal field within the sample. 
To extract the Knight shift $\mathcal{K}$, it is necessary to account for the contribution from demagnetization\cite{Xu2008}, which depends on the shape and uniform magnetization of the sample.
The crystal is approximately cylindrical, so we use the results of Ref.\ \onlinecite{JS1965}, noting that the demagnetizing field is inhomogeneous in such a non-ellipsoid, and that the sampled volume corresponds to the central region of the face of the crystal to estimate the relevant demagnetization factor $N \approx$ \num{0.739}.
Using this, we compute the Knight shift $\mathcal{K}$ as,
\begin{equation}\label{eq:demag-corr}
    \mathcal{K} = K + 4 \pi \left( N - \frac{1}{3} \right) \chi_{0}(T)
\end{equation}
where $\chi_{0}(T)$ is the CGS volume susceptibility from Ref.\ \onlinecite{Zhang2017}. 
Fig. \ref{fig:raw-shift} illustrates the effect of the demagnetization correction by comparing the raw and corrected shifts. 
In our temperature range, the susceptibility of \lno\ is almost constant, making the correction an approximately temperature independent positive offset of \SI{77+-2}{ppm}.
We note that for Li the orbital (chemical) shift $K_{\mathrm{orb}}$ is very small, at most \SI{5}{ppm}\cite{Kartha2000}, and comparable to (or less than) the uncertainty in the measurements.
We have, therefore, not attempted to account for it.

\section{Density Functional Calculations \label{App-4}}

Determination of the LNO ground state symmetry was performed using the projector augmented wave method (PAW)\cite{Blochl1994,Kresse1999} as implemented in the Vienna ab initio simulation package VASP\cite{Kresse96,Kresse1996-PRB,Paier05}.
The generalized gradient approximation (GGA) PBEsol\cite{Perdew08} was employed to account for exchange and correlation effects, while a Hubbard-like on-site repulsion $U$ and Hund's exchange coupling $J$ was introduced for the Ni-$3d$ ($U=$ \num{2.4} or \SI{6}{\electronvolt} and $J=$ \SI{1}{\electronvolt}) and La-$4f$ ($U=$ \num{14} and $J=$ \SI{1}{\electronvolt}) electrons, following the scheme of Liechtenstein {\it et al}\cite{Liechtenstein95}.
The kinetic energy cut-off was set to \SI{550}{\electronvolt} and a well-converged density of the $k$-vector mesh, corresponding to 6$\times$6$\times$6 in a simple cubic perovskite unit cell, was chosen.
We compared the total energies of the monoclinic
$P2_1/n$ unit cell of LNO and of the rhombohedral $R\bar{3}$ unit cell, both featuring two structurally inequivalent Ni sites but differing in the octahedra's rotation pattern.
It is consistently found that the rhombohedral $R\bar{3}$ unit cell has a lower energy irrespective of the magnetic order imposed [corresponding to either  $\vec{q}_{\text{AF}}=(1/2,0,0)$ or $(1/4,1/4,1/4)$ ordering vector] and the $U$ value in the GGA+$U$ scheme chosen ($U=$ \SI{6}{\electronvolt} for a robust insulating state and $U=$ \SI{2.4}{\electronvolt} for a vanishing charge gap maintaining bond disproportionation).
The energy difference between the two structures, however, is only on the order of a few meV per formula unit.

Supercell calculations were performed using both the PAW method via VASP and the linearized augmented plane wave method (LAPW) implemented in the WIEN2k code\cite{wien2k}. 
For the monoclinic $P2_1/n$ structure we used a $2a\times2a\times2a$ supercell, where $a$ is the pseudocubic lattice constant, and imposed the $\vec{q}_{\text{AF}}=(1/2,0,0)$ antiferromagnetic order onto Ni spins.
For the rhombohedral $R\bar{3}$ structure we used a face-centered $4a\times4a\times4a$ supercell and imposed the $\vec{q}_{\text{AF}}=(1/4,1/4,1/4)$ antiferromagnetic order onto Ni spins.
A single Li$^+$ ion was introduced into the various inequivalent $P$ sites (with a compensating uniform background charge) and the system was allowed to fully relax its ionic positions, with the volume fixed to the equilibrium volume of LaNiO$_3$ in PBEsol. 
The resulting total energies corresponding to different Li$^+$ positions are found to differ by no more than \SI{140}{\milli\electronvolt} per formula unit in the $P2_1/n$ structure and \SI{20}{\milli\electronvolt} per formula in the $R\bar{3}$ structure.

\section{Detailed Properties of the Interstitial \texorpdfstring{$P$}{TEXT} Site \label{App-5} }

In an attempt to explain the two \eli\ environments, here we present a detailed account of the interstitial $P$ site in the relevant pseudocubic LNO structures.
Adopting the approach used successfully in NMR of the cuprates\cite{Walstedt2008}, we focus on the hyperfine coupling to the nearest Ni atomic moments. 
A transferred hyperfine coupling will result from unpaired $3d$ ($e_g$) orbital spin density being mixed into the mostly vacant Li $2s$ orbital, where it has a Fermi contact coupling to the \eli\ nucleus.
This mixing can be due to direct overlap, or it can be mediated by the oxygen neighbours.
Symmetry considerations, analogous to the Goodenough-Kanamori rules for exchange coupling\cite{Goodenough1955,Goodenough1958,Kanamori1959}, apply to these overlaps, and we expect the coupling to the $d_{x^2 - y^2}$ orbital to be zero (both direct and through the oxygen) in the limit of a $90^\circ$ Li-O-Ni angle.
However, for all the pseudocubic distorted structures, the angles are not precisely $90^\circ$, so this coupling will be nonzero, but relatively small.
The other degenerate $e_g$ orbital, $d_{z^2}$ should also have a nonzero coupling. 
This dependence of the hyperfine coupling on angle is clearly demonstrated in the conventional NMR of transition metal oxide Li battery materials\cite{GreyRev2004}.

The Knight shift is just proportional to the spin susceptibility, $K = A \chi_s$. Assuming the macroscopic susceptibility $\chi$ is predominantly due to the Ni spins, we estimate the hyperfine coupling $A$ for the slow relaxing site as \num{0.78} kG/$\mu_B$, a value that is relatively small, e.g., compared to \eli\ in simple metals\cite{Parolin2008} or for substitutional Li in YBCO\cite{Bobroff1999}.

In the nominal rhombohedral structure of LNO, the $P$ site has two distinct Li-O-Ni angles of \SI{\sim 84}{\degree} and \SI{\sim 95}{\degree}, and two distances to the coordinating oxygens. 
However, all the $P$ sites are equivalent, so this structure is inconsistent with our two component relaxation.
Neither of the conventional lower symmetry distorted structures provide a simple doubling of the $P$ site into two equal populations of $P$-derived sites, but each inequivalent $P$ site will have a different EFG and hyperfine coupling. 
Although the multiplicity does not match the 1:1 ratio in our data, based on the above considerations, we calculated the distribution of Li-O-Ni angles for all the $P$-derived sites in both lower symmetry structures ($Pnma$ and $P2_1/n$) to see if it clustered approximately into two categories with equal weights.
We considered both the ideal structures and ones relaxed around the interstitial \lip, as estimated by DFT, but we found no indication of an appropriately symmetric bimodal distribution.

The DFT calculations, outlined in Appendix \ref{App-4}, find that the ground state structure is none of the three discussed above, but rather has symmetry $R\bar{3}$.
In this structure, the $a^{-}a^{-}a^{-}$ rotation of the NiO$_6$ octahedra of the rhombohedral unit cell ($R\bar{3}c$) is preserved; however, the symmetry is lowered by bond disproportionation.
Resulting in two Li $P$ sites (and two Ni sites), consistent with our SLR data.
However, these two sites must have a different hyperfine coupling to account for the ratio of \num{\sim 4} in SLR rates.
The distribution of Li--O--Ni angles for both $P$ sites is peaked at \SI{90}{\degree}, with one much narrower than the other.
Given the sensitivity of the hyperfine coupling to angle\cite{Pan2002}, this structure is a good candidate for explaining the two component nature of our SLR data.

\end{document}